**Spatial confinement and boundary constraints governing biological chirality: a simulation study**

Arturo Tozzi (corresponding author)
ASL Napoli 1 Centro, Distretto 27, Naples, Italy
Via Comunale del Principe 13/a 80145
tozziarturo@libero.it

ABSTRACT

Biological systems exhibit marked molecular asymmetry, with proteins based predominantly on L-amino acids and nucleic acids and carbohydrates largely composed of D-sugars. Explanations for homochirality include asymmetric photochemistry, autocatalytic amplification, stochastic symmetry breaking and mineral-surface stereoselectivity, but these mechanisms only partially address the influence of finite geometry and collective spatial interactions on stereochemical stabilization. Inspired by recent developments in condensed-matter physics, we investigated whether coherent chirality could emerge from the interplay among nonlinear stereochemical amplification, stochastic fluctuations and boundary-dependent spatial constraints. We developed a reaction-diffusion simulation in which local stereochemical populations evolved within finite two-dimensional domains under spatial coupling and weak geometrical bias fields. Our model combined bistable autocatalytic dynamics, nearest-neighbor interactions and suppression of locally inconsistent stereochemical configurations in order to quantify temporal evolution of enantiomeric excess, same-handed neighbor agreement and radial stereochemical organization under varying interaction strengths and fluctuation amplitudes. Our results showed progressive formation of chiral domains, segregation of opposite-handed regions and geometry-dependent modulation of local stereochemical organization. Spatial coupling increased local coherence and modified persistence of mixed stereochemical states, while finite boundaries influenced radial organization and anisotropic stabilization of molecular populations. Potential applications include geometrically controlled asymmetric synthesis, confined stereoselective catalytic systems, adaptive chiral materials and quantitative characterization of heterogeneous stereochemical distributions in microstructured reaction environments.

KEYWORDS: enantiomers; autocatalysis; anisotropy; microfluidics; topology.

## INTRODUCTION

Living systems exhibit a highly asymmetric molecular organization in which proteins are built predominantly from L-amino acids, whereas nucleic acids and most metabolic carbohydrates rely primarily on D-sugars (Pundir et al. 2018; Mun et al. 2019; Nshimiyimana et al. 2019; Iskandaryan et al. 2023; Hassan et al. 2025). Stereochemical selectivity propagates across multiple scales, influencing enzymatic specificity, supramolecular assembly, membrane organization and macromolecular folding (Sojo 2015; Ozturk and Sasselov 2022; Brandenburg and Hochberg 2022; Chieffo et al. 2023; Ozturk and Sasselov 2025). Explanations for homochirality include asymmetric photochemistry induced by circularly polarized radiation, parity-violating weak interactions, selective crystallization, autocatalytic amplification, stereoselective adsorption on mineral surfaces (Blackmond 2019; Blackmond 2020; Gleiser 2022; Brandenburg and Hochberg 2022; Sallembien et al. 2022; Kiliszek and Rypniewski 2023). Additional studies have examined the role of nonequilibrium thermodynamics and self-organizing reaction networks in amplifying small initial enantiomeric imbalances (Plasson et al. 2007; Michaelian 2018; Hochberg and Ribó 2019; Dyakin et al. 2021; Shi et al. 2023). These mechanisms are usually focused on microscopic chemical interactions, not fully addressing how large-scale organizational constraints may influence handedness selection and stabilization. Recent developments in condensed-matter physics have introduced related concepts in which chirality and collective order emerge from geometrical compatibility relations and constrained energetic landscapes. In layered NbOX2 crystals, transitions between achiral and chiral phases may occur through shallow energy minima connecting multiple structural states, while external perturbations like electric fields and pressure breaking symmetry degeneracy, selectively stabilizing one enantiomeric configuration over another (Gutierrez-Amigo et al. 2026). In quantum spin and dimer systems, the overall shape of the boundaries can influence how internal regions organize, even when local interaction rules remain unchanged (Shah et al. 2026). Thus, global spatial constraints may partially determine organization independently from microscopic local interactions. Together, these approaches suggest the hypothesis that compatibility relations and boundary-dependent admissibility could contribute to the emergence and stabilization of chiral states in biology.

We investigate biological homochirality as a collective spatial process shaped not only by local stereochemical amplification, but also by finite geometrical constraints and coordinated interactions among neighboring molecular populations. Our model describes interacting chiral molecular units evolving within confined energetic environments where enantiomeric states are influenced by stochastic fluctuations, short-range molecular coupling and boundary-dependent spatial organization. In our model, spatially distributed molecular populations interact through local energetic exchange, while finite boundaries regulate which large-scale stereochemical arrangements are kept stable. Inspired by the above-mentioned studies showing geometry-dependent organization and chirality selection in condensed-matter systems, our mathematical approach provides an effort to combine probabilistic transitions between stereochemical states with nonlinear autocatalytic dynamics, spatial interaction operators and suppression of locally inconsistent stereochemical configurations. Our simulations follow the temporal evolution of enantiomeric distributions under different interaction strengths, fluctuation amplitudes and geometrical conditions, aiming to identify parameter regions in which weak initial asymmetries become progressively stabilized over time.

We will proceed as follows. First, we describe the theoretical assumptions and mathematical structure of our compatibility-based stereochemical model. Then, we present the simulation design and quantitative observables to characterize enantiomeric organization under different geometrical conditions. Additionally, we distinguish organizational regimes characterized by homogeneous distributions, mixed stereochemical states or spatially segregated chiral domains. Finally, we analyze the resulting spatial and temporal patterns emerging from compatibility and boundary-dependent constraints.

## METHODS

We studied the emergence of homochirality in finite spatial domains containing interacting stereochemical populations subjected simultaneously to local autocatalytic amplification, stochastic fluctuations, compatibility-dependent spatial filtering and boundary-dependent geometrical constraints. We used a reaction-diffusion simulation in which local enantiomeric states evolved through coupled probabilistic and deterministic interactions defined on a discretized two-dimensional molecular environment. Inspired by geometry-dependent phase organization in condensed-matter systems, our model incorporated compatibility-mediated suppression of geometrically inconsistent local stereochemical configurations together with weak boundary-induced chiral bias fields. We aimed to identify experimentally discriminable signatures corresponding to stable enantiomeric segregation, spatially coherent chiral domains, persistence of compatibility-selected handedness and geometry-dependent modulation of stereochemical organization. Quantitative observables included local enantiomeric excess, same-handed neighbor agreement, mixed-state persistence and radial stereochemical profiles.

**Spatial domain**. The simulated molecular environment consisted of a finite two-dimensional spatial domain representing a prebiotic reaction surface with characteristic dimensions comparable to micrometric confined chemical systems. The computational lattice contained $N_x \times N_y$ discrete sites with spatial coordinates $(x_i, y_j)$ and lattice spacing $\Delta x = 3\ \mu m$. The total domain size was therefore $180 \times 180\ \mu m$. Spatial coordinates were defined according to

$$x_i = i\Delta x, y_j = j\Delta x$$

with $i \in [0, N_x - 1]$ and $j \in [0, N_y - 1]$. A geometrically finite elliptical reaction region was embedded inside the square computational lattice to introduce explicit boundary-dependent constraints. The admissible molecular domain was defined through the mask function

$$M(x,y) = \begin{cases} 1 & \text{if } \left(\frac{x - x_c}{R_x}\right)^2 + \left(\frac{y - y_c}{R_y}\right)^2 \leq 1 \\ 0 & \text{otherwise} \end{cases}$$

where $(x_c, y_c)$ denotes the domain center and $R_x, R_y$ correspond to the semi-major and semi-minor axes of the ellipse. Sites outside the admissible region were excluded from dynamical updates. This construction generated finite geometrical boundaries analogous to confined mesoscale systems in which bulk organization depends partially on domain geometry. Spatially resolved stereochemical states were represented by the scalar field $s(x, y, t)$, where $s = -1$ corresponded to purely D-dominant local configurations and $s = +1$ to purely L-dominant local configurations. Intermediate values represented locally mixed stereochemical populations. Initial conditions were generated according to

$$s(x, y, 0) = s_0 + \sigma_0 \eta(x, y)$$

where $s_0$ denotes the initial weak global chiral bias, $\sigma_0$ is the fluctuation amplitude and $\eta(x, y)$ is a Gaussian random variable with zero mean and unit variance. This initialization generated spatially heterogeneous stereochemical fluctuations without imposing predetermined macroscopic chirality.

**Autocatalytic dynamics**. Local stereochemical amplification was modeled through nonlinear autocatalytic dynamics analogous to bistable reaction systems capable of amplifying small initial asymmetries. The deterministic local evolution term was defined as

$$F(s) = \alpha s(1 - s^2)$$

where $\alpha$ denotes the autocatalytic amplification coefficient. This cubic formulation possesses three equilibrium points located at $s = 0$ and $s = \pm 1$. The central equilibrium corresponds to a locally racemic configuration, whereas the two outer equilibria correspond to stable chiral states. The nonlinear term $1 - s^2$ imposes saturation near strongly homochiral configurations, preventing unbounded growth. Positive local fluctuations therefore become amplified toward L-dominant states, whereas negative fluctuations evolve toward D-dominant states. The deterministic temporal evolution of the stereochemical field in the absence of spatial interactions follows

$$\frac{\partial s}{\partial t} = \alpha s(1 - s^2)$$

which generates spontaneous symmetry amplification from weak stochastic perturbations. The local free-energy structure associated with this bistable dynamics may be represented through the effective potential

$$V(s) = -\frac{\alpha}{2}s^2 + \frac{\alpha}{4}s^4$$

whose minima occur at $s = \pm 1$. This double-well structure permits local stabilization of opposite chiral configurations depending on initial fluctuations and external constraints. Temporal discretization was implemented through explicit Euler integration according to

$$s^{t+\Delta t} = s^t + \Delta t\, F(s^t)$$

with time step $\Delta t = 0.05\ h$. Numerical stability was verified by progressively reducing $\Delta t$ and confirming convergence of global stereochemical observables. Local stereochemical values were clipped to the interval $[-1,1]$ after each iteration to maintain bounded physical interpretation of enantiomeric excess.

**Spatial Coupling**. Spatial interactions between neighboring stereochemical populations were introduced through discrete diffusion-like coupling terms representing local molecular exchange and collective stereochemical alignment. The spatial coupling operator was constructed using the two-dimensional lattice Laplacian

$$\nabla^2 s_{i,j} = s_{i+1,j} + s_{i-1,j} + s_{i,j+1} + s_{i,j-1} - 4s_{i,j}$$

evaluated over nearest-neighbor sites. The corresponding diffusive interaction term was defined as

$$D\nabla^2 s$$

where $D$ denotes the effective spatial exchange coefficient measured in $h^{-1}$. This contribution promotes local smoothing of stereochemical gradients and spatial propagation of local handedness. Periodic wrapping operations were used internally for neighbor evaluation while preserving the finite elliptical mask through multiplicative spatial restriction. Outside-domain sites were assigned null stereochemical values before Laplacian evaluation to prevent artificial boundary instabilities. The full deterministic spatial evolution equation therefore became

$$\frac{\partial s}{\partial t} = \alpha s(1 - s^2) + D\nabla^2 s$$

which combines local nonlinear stereochemical amplification with spatially distributed coupling. The resulting dynamics generate spontaneous formation of coherent chiral domains separated by transition interfaces. Local interface thickness depends jointly on the diffusion coefficient $D$and the nonlinear amplification coefficient $\alpha$. The characteristic stereochemical correlation length was estimated approximately as

$$\xi \sim \sqrt{\frac{D}{\alpha}}$$

which provides a scale for the expected spatial extent of same-handed molecular clusters. Spatial heterogeneity was preserved by introducing stochastic perturbations at each temporal iteration. These fluctuations prevented immediate convergence toward globally uniform states and permitted coexistence of competing stereochemical regions.

**Compatibility filter**. Compatibility-dependent suppression of geometrically inconsistent stereochemical configurations was introduced through an additional nonlinear filtering operator inspired by compatibility-constrained collective ordering. The compatibility contribution was constructed according to

$$C(s) = \kappa\nabla^2 s(1 - \beta \mid s \mid)$$

where $\kappa$ denotes compatibility strength and $\beta$ regulates suppression near strongly homochiral states. This operator acts selectively on regions possessing large local stereochemical curvature while progressively reducing compatibility-mediated smoothing in regions already approaching stable chiral configurations. The compatibility filter therefore suppresses abrupt neighboring stereochemical inversions while preserving large coherent same-handed regions. The complete evolution equation becomes

$$\frac{\partial s}{\partial t} = \alpha s(1 - s^2) + D\nabla^2 s + \kappa\nabla^2 s(1 - \beta \mid s \mid) + \eta(x, y, t)$$

where $\eta(x, y, t)$ represents stochastic fluctuations. The compatibility term modifies the effective propagation of local stereochemical interfaces by penalizing geometrically inconsistent local configurations. The resulting evolution selectively stabilizes collective stereochemical arrangements compatible with neighboring spatial organization. The compatibility coefficient $\kappa$ was systematically varied between 0 and 0.55 $h^{-1}$ during parameter sensitivity analyses. Increasing $\kappa$ progressively reduced persistence of fragmented local stereochemical patches and promoted formation of larger spatially coherent domains. Numerical evaluation of compatibility effects was implemented at each iteration following computation of the local Laplacian operator. Compatibility-dependent updates were applied before stochastic perturbations to preserve deterministic filtering of local stereochemical inconsistencies.

**Boundary bias**. Boundary-dependent geometrical constraints were incorporated through a weak spatial bias field localized preferentially near finite domain boundaries. The normalized radial coordinate was defined as

$$\rho(x, y) = \sqrt{\left(\frac{x - x_c}{R_x}\right)^2 + \left(\frac{y - y_c}{R_y}\right)^2}$$

such that $\rho = 1$ corresponds approximately to the elliptical boundary. The boundary enhancement function was then defined according to

$$B(\rho) = \max\left(0, \frac{\rho - \rho_c}{1 - \rho_c}\right)$$

where $\rho_c$ specifies the onset radius of boundary amplification. The resulting stereochemical bias field was implemented as

$$\Phi(x, y) = \gamma B(\rho) \cos(\theta - \theta_0)$$

where $\gamma$ is the boundary bias amplitude, $\theta$ denotes the local angular coordinate and $\theta_0$ specifies the preferred directional orientation. This weak anisotropic modulation introduces geometry-dependent local asymmetry analogous to externally biased chirality selection. The full update equation therefore becomes

$$s^{t+\Delta t} = s^t + \Delta t[F(s) + D\nabla^2 s + C(s) + \Phi(x, y)] + \sqrt{\Delta t}\,\sigma\eta(x, y, t)$$

where $\sigma$ denotes fluctuation amplitude. Boundary bias amplitudes remained substantially smaller than the nonlinear amplification coefficient to avoid externally imposing global chirality. Instead, the bias field acted as a weak geometrical perturbation capable of influencing stereochemical organization preferentially near finite boundaries. Spatial radial profiles of local enantiomeric excess were subsequently computed to evaluate geometry-dependent modulation of stereochemical organization across the domain.

**Quantitative measures**. Several quantitative observables were computed throughout the simulations to characterize stereochemical organization. The global enantiomeric excess was calculated according to

$$EE(t) = \frac{1}{N}\sum_{i,j} s_{i,j}(t)$$

where $N$ denotes the number of admissible spatial sites. This quantity measures the net population-level stereochemical asymmetry. Local mixed-state persistence was evaluated through

$$M(t) = \frac{1}{N}\sum_{i,j} H(\epsilon - \mid s_{i,j}(t) \mid)$$

where $H$ is the Heaviside function and $\epsilon = 0.20$ defines the threshold separating weakly chiral from strongly chiral local states. Spatial stereochemical coherence was quantified using the same-handed neighbor agreement index

$$A(t) = \frac{1}{N_n}\sum_{\langle i,j \rangle} \delta\left[\text{sign}(s_i), \text{sign}(s_j)\right]$$

where $N_n$ denotes the number of neighboring site pairs and $\delta$ is the Kronecker delta. This observable measures local spatial coherence of stereochemical organization. Radial stereochemical profiles were computed through annular averaging:

$$R(r) = \frac{1}{N_r}\sum_{(x,y)\in r} s(x, y)$$

where $N_r$ denotes the number of sites inside the radial shell centered at distance $r$. Histograms of local stereochemical states were estimated using normalized density distributions over the interval $[-1,1]$. Parameter sensitivity analyses were performed by systematically varying compatibility strength and fluctuation amplitude while maintaining identical initial geometrical conditions. Each simulation was executed for $36\ h$ of simulated temporal evolution corresponding to $720$ numerical iterations. Random number generation used pseudo-random Gaussian sampling implemented through NumPy random generators. Numerical integration, matrix operations, array manipulations and graphical rendering were performed using Python 3.11 together with NumPy 2.0 and Matplotlib 3.9._

RESULTS

We report the emergence of compatibility-dependent stereochemical organization in finite molecular domains subjected to local autocatalytic amplification, stochastic fluctuations and geometrical boundary constraints. Quantitative analyses examined the temporal evolution of enantiomeric excess, spatial segregation, same-handed neighbor agreement and radial stereochemical organization across different compatibility strengths and fluctuation amplitudes. The simulations revealed progressive formation of coherent chiral domains together with geometry-dependent modulation of spatial stereochemical organization.

**Spatial evolution**. Simulations initialized with weak stereochemical asymmetry evolved toward progressively organized chiral domains characterized by spatial segregation of locally L-dominant and D-dominant regions (Figure 1). Early temporal stages between 0 h and 6 h displayed heterogeneous local fluctuations distributed throughout the elliptical molecular domain without persistent large-scale ordering (Figure 1, upper-row left and upper-row second insets). Between 18 h and 36 h, compatibility-constrained simulations progressively generated extended same-handed territories separated by narrow transition interfaces (Figure 1, upper-row third and upper-row right insets). The final local enantiomeric excess distributions became bimodal, indicating accumulation near strongly chiral states and progressive depletion of locally mixed configurations (Figure 1, lower-right inset). Temporal trajectories of the global enantiomeric excess showed divergence between simulations with and without compatibility filtering (Figure 1, second-row left inset). Across 20 independent realizations, the mean absolute final enantiomeric excess increased from $0.267 \pm 0.109$ in unconstrained simulations to $0.418 \pm 0.103$ under compatibility filtering (Welch two-sample testing: $t = 4.97$, $p = 1.05 \times 10^{-5}$). Spatial coherence similarly increased under compatibility constraints. The same-handed neighbor agreement index reached $0.974 \pm 0.006$ in compatibility-constrained simulations compared with $0.932 \pm 0.019$ in unconstrained conditions ($t = 11.8$, $p = 1.67 \times 10^{-15}$). Local mixed-state persistence remained low throughout the simulations but showed modest increases under compatibility filtering, with final mixed fractions of $0.020 \pm 0.006$ compared with $0.013 \pm 0.005$in unconstrained domains ($t = 4.59$, $p = 4.09 \times 10^{-5}$) (Figure 1, lower-left inset). Simulated concentrations of L-like and D-like molecular pools progressively diverged during temporal evolution, consistent with stabilization of coherent stereochemical asymmetry (Figure 1, second-row right inset).
These results suggest that compatibility-dependent interactions influenced both the extent and spatial distribution of stereochemical asymmetry, promoting coherent chiral organization while maintaining localized transition zones between neighboring opposite-handed regions.

**Geometry effects**. Parameter sensitivity analyses showed that stereochemical organization depended jointly on compatibility strength, fluctuation amplitude and finite-domain geometry (Figure 2). Increasing compatibility coefficients progressively shifted the simulations toward larger same-handed spatial domains and higher final enantiomeric excess values, whereas increasing stochastic fluctuation amplitudes reduced persistence of coherent stereochemical organization (Figure 2, upper-left and upper-middle insets). Heatmap analyses showed that organized chiral states emerged preferentially within intermediate fluctuation regimes, where stochastic perturbations remained sufficiently strong to generate local asymmetry while not overwhelming compatibility-mediated spatial stabilization. Correlation analysis between same-handed spatial agreement and final enantiomeric excess across the explored parameter space produced a Pearson coefficient of $r = 0.71$, indicating partial coupling between local spatial coherence and global stereochemical asymmetry (Figure 2, upper-right inset). Radial stereochemical profiles additionally demonstrated geometry-dependent modulation of local chirality. Absolute enantiomeric excess increased progressively from the central region toward boundary-adjacent sectors, pointing towards enhanced stabilization of coherent stereochemical states near regions exposed to boundary-dependent bias fields (Figure 2, lower-left inset). Spatial maps of the compatibility-modulated bias term showed anisotropic geometrical influence concentrated preferentially along peripheral sectors of the elliptical domain (Figure 2, lower-middle inset). Final stereochemical configurations therefore exhibited coexistence of extended coherent regions together with localized mixed interfaces, particularly near geometrically constrained transition zones (Figure 2, lower-right inset). Histograms of final local stereochemical states revealed progressive redistribution away from weakly chiral configurations toward strongly polarized local domains as compatibility strength increased.
Collectively, these observations indicate that geometrical admissibility and compatibility filtering jointly regulated the persistence, segregation and spatial coherence of stereochemical populations within finite molecular environments.

Overall, our results suggest that local autocatalytic amplification alone did not fully determine the spatial organization of stereochemical asymmetry. Compatibility-dependent interactions and finite geometrical boundaries modified the emergence and stabilization of collective chiral states by influencing spatial coherence, interface persistence and radial stereochemical organization. Therefore, our simulations support the hypothesis that biological homochirality may depend not exclusively on local chemical amplification, but also on compatibility-mediated collective organization within geometrically constrained environments.

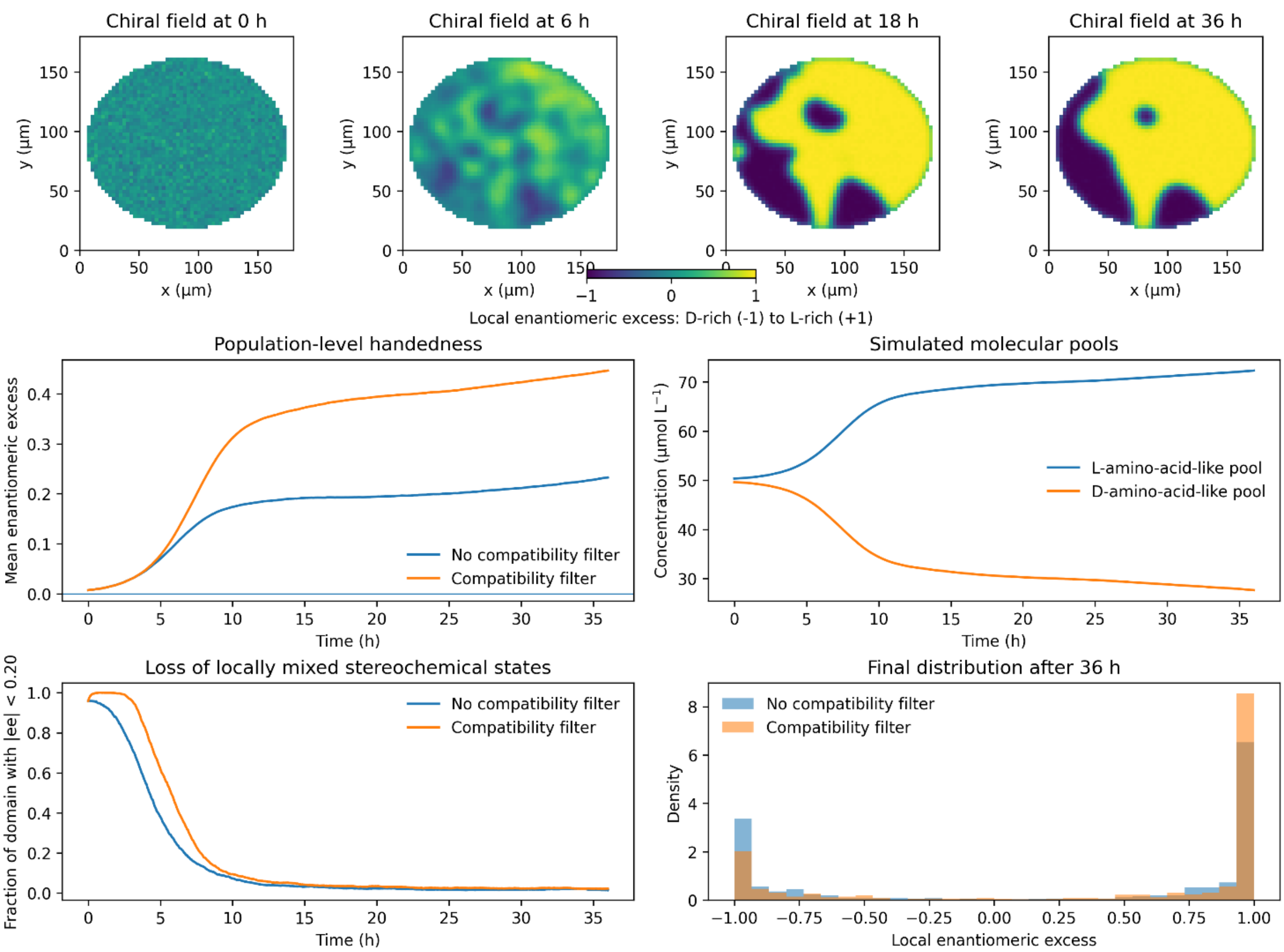

**Figure 1.** Compatibility-constrained emergence of stereochemical organization within a finite molecular domain. Insets illustrate the simulated temporal evolution of interacting chiral molecular populations inside a two-dimensional reaction environment measuring $180 \times 180\ \mu m$, showing the progressive formation, stabilization and spatial segregation of coherent stereochemical regions over time.

Upper row: spatial maps of local enantiomeric excess at 0 h, 6 h, 18 h and 36 h. Early stages exhibit mixed local stereochemistry with small fluctuations distributed across the domain. Progressive evolution generates spatial segregation and the formation of coherent chiral territories.

Second-row left inset: temporal evolution of the mean enantiomeric excess for simulations with and without compatibility filtering. The compatibility-constrained system develops a persistent global stereochemical bias, whereas the unconstrained condition remains closer to fluctuating mixed states.

Second-row right inset: the corresponding concentrations of simulated L-like and D-like molecular pools expressed in μmol $L^{-1}$.

Lower-left inset: persistence of weakly asymmetric stereochemical regions over time.

Lower-right inset: the final probability distributions of local stereochemical states after 36 h, showing broader segregation and increased accumulation near strongly chiral configurations under compatibility constraints.

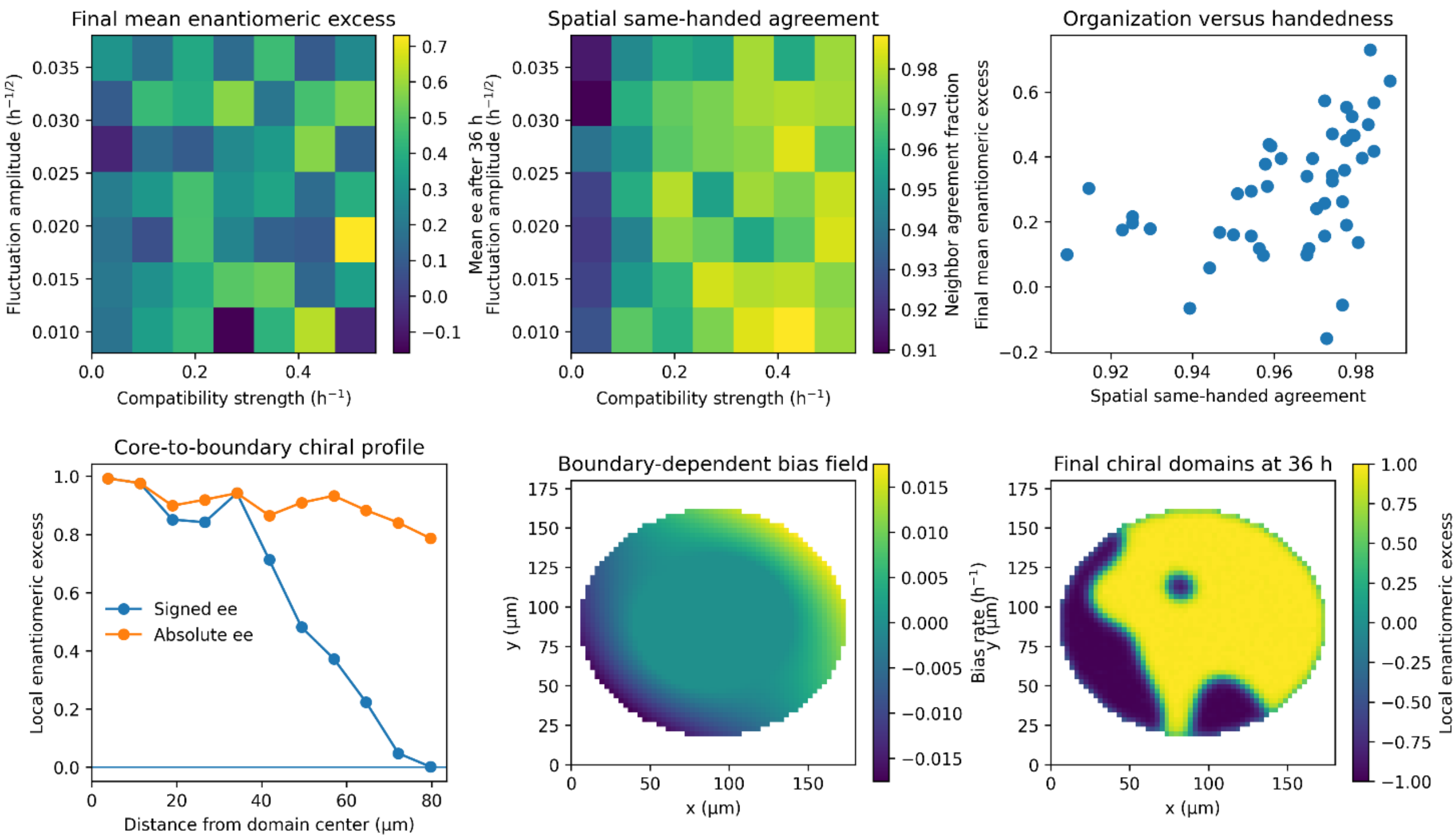

**Figure 2.** Geometry-dependent collective stereochemical organization under varying compatibility and fluctuation conditions.
Upper-left heatmap: final mean enantiomeric excess measured after 36 h across different compatibility strengths and stochastic fluctuation amplitudes.
Upper-middle heatmap: the corresponding same-handed spatial agreement index, quantifying local stereochemical coherence among neighboring regions.
Upper-right inset: relationship between spatial organization and global handedness, indicating increased stereochemical asymmetry in simulations with stronger same-handed spatial correlations.
Lower-left inset: radial profile of local enantiomeric excess from the center toward the domain boundary, revealing spatial heterogeneity associated with geometrical constraints.
Lower-middle inset: boundary-dependent compatibility bias field applied to the simulated domain, expressed as a local bias rate in $h^{-1}$.
Lower-right inset: the final spatial organization of chiral domains after 36 h, highlighting the coexistence of spatially distinct stereochemical regions shaped by compatibility interactions and finite geometrical boundaries.

## CONCLUSIONS

We asked whether biological homochirality could emerge not exclusively from local stereochemical amplification mechanisms, but also from collective compatibility relations and finite geometrical constraints capable of influencing the spatial stabilization of chiral molecular populations. Our simulations compared unconstrained stereochemical dynamics with compatibility-dependent organization in finite two-dimensional molecular environments subjected to stochastic fluctuations, nonlinear autocatalytic amplification and boundary-modulated spatial interactions. We found that compatibility-dependent coupling altered both the magnitude and spatial coherence of enantiomeric organization, generating extended same-handed molecular regions together with localized transition interfaces separating neighboring stereochemical domains. Simulations also showed that boundary geometry modulated radial stereochemical organization and influenced the persistence of coherent chiral territories. Parameter sensitivity analyses indicated that compatibility strength and fluctuation amplitude jointly regulated the emergence of organized stereochemical states, with intermediate fluctuation regimes favoring stabilization of collective asymmetry. These observations are consistent with the hypothesis that homochiral organization may depend on geometrical admissibility and collective spatial interactions in addition to local chemical amplification. Our results support the view that stereochemical asymmetry can be examined as a multiscale organizational process involving both local reaction dynamics and finite-domain collective constraints.

Compared with autocatalytic amplification models, racemization kinetics and purely reaction-diffusion descriptions of homochirality (Stepensky et al. 2004; Causin and Facchetti 2009; Ott and Topczewski 2018; Blackmond 2020; Kim et al. 2021; Higgs and Blackmond 2025; Pirovino et al. 2025), our approach incorporated compatibility-dependent suppression of spatially inconsistent stereochemical configurations together with geometrically finite domains. Existing techniques describe chirality emergence primarily through local nonlinear reaction terms, asymmetric catalysis, stochastic symmetry breaking or mineral-surface stereoselectivity, without introducing collective spatial admissibility conditions acting across extended molecular populations (Satyanarayana et al. 2009; Knutson et al. 2022; Jiang and Lewis 2023; Ozturk et al. 2023; Xiao and Zhao 2023). In contrast, our simulations coupled bistable stereochemical dynamics with compatibility-mediated filtering operators and boundary-dependent anisotropic modulation. Unlike mean-field kinetic descriptions that neglect mesoscale spatial organization, our model preserved local stereochemical heterogeneity, interface formation and finite-domain geometry throughout temporal evolution. Our simulations integrated also quantitative measures of neighbor agreement, radial stereochemical organization and local mixed-state persistence, allowing simultaneous characterization of global asymmetry and spatial coherence.

Our approach has limitations. Simulations should be interpreted as representative conceptual models rather than experimentally validated biochemical systems. The compatibility operator does not derive from an established stereochemical law equivalent to compatibility relations in elasticity theory. Consequently, the relationship between compatibility-dependent filtering and actual molecular chirality selection is conceptual. The bistable autocatalytic dynamics employed here are mathematically standard, but the connection between mesoscale spatial organization and biochemical homochirality is indirect. The imposed boundary bias field introduces externally specified anisotropy, meaning that geometry contributes partially through predefined directional modulation rather than emerging entirely spontaneously from molecular interactions. Statistical significance reflects properties of the selected equations and parameters rather than empirical biological measurements. Concentration scales and spatial dimensions were chosen to preserve physical interpretability, but were not calibrated against specific prebiotic environments. Aso, spatial segregation patterns may appear sharper than those expected in realistic prebiotic chemistry because our simulations do not incorporate heterogeneous diffusion coefficients, reaction intermediates, catalytic degradation pathways or multicomponent kinetic coupling. Still, the analogies with condensed-matter systems are heuristic.

Testable hypotheses can be suggested. First, finite geometrical confinement modifies the temporal stabilization of enantiomeric excess in autocatalytic stereochemical systems. Microfluidic reaction chambers with identical chemical composition but different boundary geometries could therefore display distinct spatial chirality distributions. Our simulations predict that increasing geometrical anisotropy should increase same-handed spatial agreement by approximately 3–5% relative to isotropic domains under comparable fluctuation amplitudes.
Second, weak externally imposed directional perturbations near system boundaries alter radial stereochemical organization without necessarily imposing complete global homochirality. This prediction may be tested through localized electric-field exposure or surface-patterned catalytic substrates, with measurable observables including radial enantiomeric excess profiles and interface persistence times.
Third, intermediate stochastic fluctuation regimes maximize stable spatial segregation of chiral domains. Our simulations predict a nonmonotonic dependence between fluctuation amplitude and spatial coherence, with maximal organization occurring within bounded fluctuation intervals rather than under purely deterministic conditions.
Another experimentally measurable prediction concerns interface density between opposite-handed regions. Increasing compatibility strength in the simulations reduced fragmented local domains and increased coherent cluster size, suggesting that experimentally measured stereochemical interface density should scale inversely with collective coupling strength.
Unanswered questions include how multiscale reaction networks modify spatial chirality propagation and whether experimentally measurable molecular systems exhibit analogous geometry-dependent stabilization of collective stereochemical states. Future research may incorporate heterogeneous diffusion tensors, multicomponent catalytic networks, explicit reaction intermediates and chemically realistic kinetic constants.

Potential practical applications include spatially controlled stereoselective synthesis in confined reaction environments, optimization of asymmetric catalytic microreactors and geometrically regulated production of chiral molecular assemblies. Compatibility-dependent stereochemical organization may contribute to the design of adaptive reaction surfaces capable of stabilizing spatially coherent enantiomeric distributions under fluctuating environmental conditions. Geometrically constrained molecular systems could support development of tunable chiral materials whose stereochemical organization depends on finite-domain morphology rather than exclusively on molecular composition. Quantitative descriptors derived from spatial chirality organization may assist characterization of heterogeneous stereochemical distributions in confined biochemical reactors, catalytic membranes and synthetic protocell systems. Further, controlled modulation of local stereochemical coherence could contribute to nanoscale assembly processes involving helicoidal polymers, supramolecular aggregates or anisotropic molecular scaffolds.

In conclusion, we investigated how collective compatibility relations and finite geometrical constraints could influence the emergence of organized stereochemical asymmetry in spatially extended molecular systems. Our simulations identified coupled effects linking local nonlinear amplification, stochastic perturbations and geometry-dependent collective organization. Our observations support the interpretation that stereochemical order may depend simultaneously on microscopic reaction dynamics and mesoscale spatial admissibility conditions acting across finite molecular populations.

## DECLARATIONS

**Ethics approval and consent to participate.** This research does not contain any studies with human participants or animals performed by the Author.
**Consent for publication.** The Author transfers all copyright ownership, in the event the work is published. The undersigned author warrants that the article is original, does not infringe on any copyright or other proprietary right of any third part, is not under consideration by another journal and has not been previously published.
**Availability of data and materials.** All data and materials generated or analyzed during this study are included in the manuscript. The Author had full access to all the data in the study and took responsibility for the integrity of the data and the accuracy of the data analysis.
**Disclaimer**. The views expressed are those of the author and do not necessarily reflect those of the affiliated institutions.
**Competing interests.** The Author does not have any known or potential conflict of interest including any financial, personal or other relationships with other people or organizations within three years of beginning the submitted work that could inappropriately influence or be perceived to influence their work.
**Funding.** This research did not receive any specific grant from funding agencies in the public, commercial or not-for-profit sectors.
**Acknowledgements:** none.
**Authors' contributions.** The Author performed: study concept and design, acquisition of data, analysis and interpretation of data, drafting of the manuscript, critical revision of the manuscript for important intellectual content, statistical analysis, obtained funding, administrative, technical and material support, study supervision.
**Declaration of generative AI and AI-assisted technologies in the writing process.** During the preparation of this work, the author used ChatGPT 5.3 to assist with data analysis and manuscript drafting and to improve spelling, grammar and general editing. After using this tool, the author reviewed and edited the content as needed, taking full responsibility for the content of the publication.

## REFERENCES

1) Blackmond, D. G. 2019. "The Origin of Biological Homochirality." *Cold Spring Harbor Perspectives in Biology* 11 (3): a032540. https://doi.org/10.1101/cshperspect.a032540.
2) Blackmond, D. G. 2020. "Autocatalytic Models for the Origin of Biological Homochirality." *Chemical Reviews* 120 (11): 4831–4847. https://doi.org/10.1021/acs.chemrev.9b00557.
3) Blackmond, D. G. 2020. "Autocatalytic Models for the Origin of Biological Homochirality." *Chemical Reviews* 120 (11): 4831–4847. https://doi.org/10.1021/acs.chemrev.9b00557.
4) Brandenburg, A., and D. Hochberg. 2022. "Introduction to Origins of Biological Homochirality." *Origins of Life and Evolution of Biospheres* 52 (1–3): 1–2. https://doi.org/10.1007/s11084-022-09629-4.
5) Brandenburg, A., and D. Hochberg. 2022. "Introduction to Origins of Biological Homochirality." *Origins of Life and Evolution of Biospheres* 52 (1–3): 1–2. https://doi.org/10.1007/s11084-022-09629-4.
6) Causin, P., and G. Facchetti. 2009. "Autocatalytic Loop, Amplification and Diffusion: A Mathematical and Computational Model of Cell Polarization in Neural Chemotaxis." *PLoS Computational Biology* 5 (8): e1000479. https://doi.org/10.1371/journal.pcbi.1000479.
7) Chieffo, C., A. Shvetsova, F. Skorda, A. Lopez, and M. Fiore. 2023. "The Origin and Early Evolution of Life: Homochirality Emergence in Prebiotic Environments." *Astrobiology* 23 (12): 1368–1382. https://doi.org/10.1089/ast.2023.0007.
8) Dyakin, V. V., N. V. Dyakina-Fagnano, L. B. McIntire, and V. N. Uversky. 2021. "Fundamental Clock of Biological Aging: Convergence of Molecular, Neurodegenerative, Cognitive and Psychiatric Pathways: Non-Equilibrium Thermodynamics Meet Psychology." *International Journal of Molecular Sciences* 23 (1): 285. https://doi.org/10.3390/ijms23010285.
9) Gleiser, M. 2022. "Biological Homochirality and the Search for Extraterrestrial Biosignatures." *Origins of Life and Evolution of Biospheres* 52 (1–3): 93–104. https://doi.org/10.1007/s11084-022-09623-w.
10) Gutierrez-Amigo, Martin, Claudia Felser, Ion Errea, and Maia G. Vergniory. 2026. "Emergent Chirality and Enantiomeric Selectivity in Layered NbOX2 Crystals." *Physical Review Letters* 136: 166605. https://doi.org/10.1103/kb6r-zxwq.
11) Hassan, A., R. di Vito, T. Nuzzo, M. Vidali, M. J. Carlini, S. Yadav, H. Yang, A. D'Amico, X. Kolici, V. Valsecchi, C. Panicucci, G. Pignataro, C. Bruno, E. Bertini, F. Errico, L. Pellizzoni, and A. Usiello. 2025. "Dysregulated Balance of D- and L-Amino Acids Modulating Glutamatergic Neurotransmission in Severe Spinal Muscular Atrophy." *Neurobiology of Disease* 207: 106849. https://doi.org/10.1016/j.nbd.2025.106849.
12) Higgs, P. G., and D. G. Blackmond. 2025. "Autocatalytic Symmetry Breaking and Chiral Amplification in a Feedback Network Combining Amino Acid Synthesis and Ligation." *Proceedings of the National Academy of Sciences of the United States of America* 122 (20): e2423683122. https://doi.org/10.1073/pnas.2423683122.

13) Hochberg, D., and J. M. Ribó. 2019. "Entropic Analysis of Mirror Symmetry Breaking in Chiral Hypercycles." *Life* 9 (1): 28. https://doi.org/10.3390/life9010028.
14) Iskandaryan, M., S. Blbulyan, M. Sahakyan, A. Vassilian, K. Trchounian, and A. Poladyan. 2023. "L-Amino Acids Affect the Hydrogenase Activity and Growth of *Ralstonia eutropha* H16." *AMB Express* 13 (1): 33. https://doi.org/10.1186/s13568-023-01535-w.
15) Jiang, Y., and J. C. Lewis. 2023. "Asymmetric Catalysis by Flavin-Dependent Halogenases." *Chirality* 35 (8): 452–460. https://doi.org/10.1002/chir.23550.
16) Kiliszek, A., and W. Rypniewski. 2023. "The Emergence of Biological Homochirality." *Acta Biochimica Polonica* 70 (3): 481–485. https://doi.org/10.18388/abp.2020_6914.
17) Kim, S., A. Martínez Dibildox, A. Aguirre-Soto, and H. D. Sikes. 2021. "Exponential Amplification Using Photoredox Autocatalysis." *Journal of the American Chemical Society* 143 (30): 11544–11553. https://doi.org/10.1021/jacs.1c04236.
18) Knutson, P. C., H. Ji, C. M. Harrington, Y. T. Ke, and E. M. Ferreira. 2022. "Chirality Transfer and Asymmetric Catalysis: Two Strategies toward the Enantioselective Formal Total Synthesis of (+)-Gelsenicine." *Organic Letters* 24 (27): 4971–4976. https://doi.org/10.1021/acs.orglett.2c01974.
19) Michaelian, K. 2018. "Homochirality through Photon-Induced Denaturing of RNA/DNA at the Origin of Life." *Life* 8 (2): 21. https://doi.org/10.3390/life8020021.
20) Mun, H. C., K. M. Leach, and A. D. Conigrave. 2019. "L-Amino Acids Promote Calcitonin Release via a Calcium-Sensing Receptor: Gq/11-Mediated Pathway in Human C-Cells." *Endocrinology* 160 (7): 1590–1599. https://doi.org/10.1210/en.2018-00860.
21) Nshimiyimana, P., L. Liu, and G. Du. 2019. "Engineering of L-Amino Acid Deaminases for the Production of α-Keto Acids from L-Amino Acids." *Bioengineered* 10 (1): 43–51. https://doi.org/10.1080/21655979.2019.1595990.
22) Ott, A. A., and J. J. Topczewski. 2018. "Catalytic Racemization of Activated Organic Azides." *Organic Letters* 20 (22): 7253–7256. https://doi.org/10.1021/acs.orglett.8b03168.
23) Ozturk, S. F., and D. D. Sasselov. 2022. "On the Origins of Life's Homochirality: Inducing Enantiomeric Excess with Spin-Polarized Electrons." *Proceedings of the National Academy of Sciences of the United States of America* 119 (28): e2204765119. https://doi.org/10.1073/pnas.2204765119.
24) Ozturk, S. F., D. D. Sasselov, and J. D. Sutherland. 2023. "The Central Dogma of Biological Homochirality: How Does Chiral Information Propagate in a Prebiotic Network?" *Journal of Chemical Physics* 159 (6): 061102. https://doi.org/10.1063/5.0156527.
25) Ozturk, S. F., and D. D. Sasselov. 2025. "Life's Homochirality: Across a Prebiotic Network." *Proceedings of the National Academy of Sciences of the United States of America* 122 (34): e2505126122. https://doi.org/10.1073/pnas.2505126122.
26) Pirovino, M., C. Iseli, J. A. Curran, and B. Conrad. 2025. "Biomathematical Enzyme Kinetics Model of Prebiotic Autocatalytic RNA Networks: Degenerating Parasite-Specific Hyperparasite Catalysts Confer Parasite Resistance and Herald the Birth of Molecular Immunity." *PLoS Computational Biology* 21 (1): e1012162. https://doi.org/10.1371/journal.pcbi.1012162.
27) Plasson, R., D. K. Kondepudi, H. Bersini, A. Commeyras, and K. Asakura. 2007. "Emergence of Homochirality in Far-from-Equilibrium Systems: Mechanisms and Role in Prebiotic Chemistry." *Chirality* 19 (8): 589–600. https://doi.org/10.1002/chir.20440.
28) Pundir, C. S., S. Lata, and V. Narwal. 2018. "Biosensors for Determination of D and L-Amino Acids: A Review." *Biosensors and Bioelectronics* 117: 373–384. https://doi.org/10.1016/j.bios.2018.06.033.
29) Sallembien, Q., L. Bouteiller, J. Crassous, and M. Raynal. 2022. "Possible Chemical and Physical Scenarios Towards Biological Homochirality." *Chemical Society Reviews* 51 (9): 3436–3476. https://doi.org/10.1039/d1cs01179k.
30) Satyanarayana, T., S. Abraham, and H. B. Kagan. 2009. "Nonlinear Effects in Asymmetric Catalysis." *Angewandte Chemie International Edition* 48 (3): 456–494. https://doi.org/10.1002/anie.200705241.
31) Shah, Jeet, Laura Shou, Jeremy Shuler, and Victor Galitski. 2026. "Breakdown of the Thermodynamic Limit in Quantum Spin and Dimer Models." *Physical Review X* 16: 021020. https://doi.org/10.1103/ckrx-wbct.
32) Shi, W., K. Liang, R. Wang, J. Liu, and C. Lu. 2023. "Biased Symmetry Breaking in the Formation of Intercalated Layered Double Hydroxides: Toward Control of Homochiral Supramolecular Assembly." *Small* 19 (44): e2303497. https://doi.org/10.1002/smll.202303497.
33) Sojo, V. 2015. "On the Biogenic Origins of Homochirality." *Origins of Life and Evolution of Biospheres* 45 (1–2): 219–224. https://doi.org/10.1007/s11084-015-9422-9.
34) Stepensky, D., M. Chorny, Z. Dabour, and I. Schumacher. 2004. "Long-Term Stability Study of L-Adrenaline Injections: Kinetics of Sulfonation and Racemization Pathways of Drug Degradation." *Journal of Pharmaceutical Sciences* 93 (4): 969–980. https://doi.org/10.1002/jps.20010.
35) Xiao, X., and B. Zhao. 2023. "Vitamin B6-Based Biomimetic Asymmetric Catalysis." *Accounts of Chemical Research* 56 (9): 1097–1117. https://doi.org/10.1021/acs.accounts.3c00053.